\documentclass[aps,preprint,nofootinbib,floatfix]{revtex4}
\usepackage{amsmath}
\usepackage{epsfig}

\input{tcilatex}

\begin{document}

\title{$SO(2,C)$ Invariant Discrete Gauge States in Liouville Gravity
Coupled to Minimal Conformal Matter}
\author{Jen-Chi Lee}
\affiliation{Department of Electrophysics, National Chiao-Tung University, Hsinchu,
Taiwan 30050, R.O.C. (e-mail:jcclee@cc.nctu.edu.tw)}
\date{27 March 1997}

\begin{abstract}
We contruct the general formula for a set of discrete gauge states (DGS) in $%
c<1$ Liouville theory. This formula reproduces the previously found $c=1$
DGS in the appropriate limiting case. We also demonstrate the $SO(2,C)$
invariant structure of these DGS in the old covariant quantization of the
theory. This is in analogy to the $SO(2,C)$ invariant ring structure of BRST
cohomology of the theory.
\end{abstract}

\maketitle

\address{Department of Electrophysics, National Chiao Tung University,
Hsinchu 300, Taiwan}





\section{Introduction}

Liouville gravity \cite{I.R Klebanov} has been an important consistency
check of the discretized matrix model approach to non-perturbative string
theory for the last few years. Many interesting phenomena, which were
peculiar to 2D string, were uncovered and attempts have been made to compare
these results to the more realistic high dimensional string theory \cite%
{J.C. Lee}. There are two quantization schemes of the theory which appeared
in the literature. In the most popular BRST approach, the ground ring
structure of ghost number zero operators accounts for the existence of
space-time $w_{\infty }$ symmetry of the theory \cite{I.R. Klebanov}. In the
old covariant quantization (OCQ) scheme, we believe that the gauge states 
\cite{T.D. Chung1} (physical zero-norn states) \ will play the role of
nontrivial ghost sector in BRST approach. However, unlike the discrete
Polyakov states, there is an infinite number of continuum momentum gauge
states in the massive levels of the spectrum, and it is difficult to give a
general formula for them. In previous papers \cite{J.C. Lee}\cite{T.D.
Chung1}, we introduced the concept of DGS (gauge states with Polyakov
discrete momentum) in the OCQ scheme and explicitly constructed a set of
them for $c=1$ Liouville theory. We then showed that they do carry the $%
w_{\infty }$\ charges. An explicit form of worldsheet supersymmetric DGS in
\ $N=1$ super Lioville theory was also given in \cite{T.D. Chung2}.

Since the idea of gauge states has direct application in the high
dimensional string theory \cite{J.C. Lee}, it would be important to
understand it more in general Liouville theory and compare it with the known
results in BRST Liouville theory. In this paper, we will show that the $%
SO(2,C)$ invariant ring structure of BRST cohomology of the $c\leq 1$
Liouville model \cite{B.Lian} has its counterpart in OCQ scheme, that is $the
$ $SO(2,C)$ $invariant$ DGS. We will first construct the general formula for
a set of DGS in $c<1$ Liouville theory. This formula reduces to the \ $c=1$
DGS in the $p=q+1,q\rightarrow \infty $ limit. We then show that the
particular set of DGS we constructed together with the c = 1 DGS discovered
previously \cite{T.D. Chung1} form a $SO(2,C)$ invariant set. It is thus
easily seen that this $c<1$ DGS also carrys the $w_{\infty }$ charges.

\section{Discrete gauge states in c\TEXTsymbol{<}1 Liouville theory}

We consider the following action of $c<1$ Liouville theory \cite{Ken-Ji}%
\begin{eqnarray}
S &=&\frac{1}{8\pi }\int d^{2}z\sqrt{\widehat{g}}[\widehat{g}^{\alpha \beta
}\partial _{\alpha }X\partial _{\beta }X+2iQ_{M}\widehat{R}X  \notag \\
&&+\widehat{g}^{\alpha \beta }\partial _{\alpha }\phi \partial _{\beta }\phi
+2Q_{L}\widehat{R}\phi ],  \TCItag{2.1}
\end{eqnarray}%
with $\phi $ being the Liouville field. If we take 
\begin{equation}
Q_{M}=(p-q)Q,Q_{L}=(p+q)Q  \tag{2.2}
\end{equation}%
with $Q=\frac{1}{\sqrt{2pq}}$ and p,q are two coprime positive integers, the
central charges for both fields will be 
\begin{equation}
c_{M}=1-\frac{6(p-q)^{2}}{pq},c_{L}=1-\frac{6(p+q)^{2}}{pq}  \tag{2.3}
\end{equation}%
so that 
\begin{equation}
c_{M+}c_{L}=26,  \tag{2.4}  \label{CMCL}
\end{equation}%
which cancels the anomaly from ghost contribution. The stress energy tensor
is ( from now on we consider the chiral sector only) 
\begin{equation}
T_{zz}=-\frac{1}{2}(\partial X)^{2}+iQ_{M}\partial ^{2}X-\frac{1}{2}%
(\partial \phi )^{2}-Q_{L}\partial ^{2}\phi .  \tag{2.5}
\end{equation}%
Note that if we take the $p=q+1,q\rightarrow \infty \ $limit, we recover the
usual unitary $c_{M}=1$ \ 2D gravity model. The mode expansion of $X^{\mu
}(\phi ,X)$ is defined to be%
\begin{equation}
\partial _{z}X^{\mu }=-\overset{\infty }{\underset{-\infty }{\sum }}%
z^{-n-1}(\alpha _{n}^{0},i\alpha _{n}^{1}),  \tag{2.6}
\end{equation}%
\bigskip with the metric $\eta _{\mu \nu }=diag[-1,1],$ $Q^{\mu
}=(iQ_{L},Q_{M})$ and the zero mode $\alpha _{0}^{\mu }=f^{\mu }=(\epsilon
,p)$. The corresponding Virasoro generators are 
\begin{eqnarray}
L_{n} &=&(\frac{n+1}{2}Q^{\mu }+f^{\mu })\alpha _{\mu ,n}  \notag \\
+\frac{1}{2}\underset{k\neq 0}{\sum } &:&\alpha _{\mu ,-k}\alpha _{n+k}^{\mu
}:n\neq 0,  \TCItag{2.7}
\end{eqnarray}

\begin{equation}
L_{0}=\frac{1}{2}(Q^{\mu }+f^{\mu })f^{\mu }+\underset{k=1}{\overset{\infty }%
{\sum }}:\alpha _{\mu ,-k}\alpha _{k}^{\mu }:.  \tag{2.8}
\end{equation}%
In the OCQ scheme, physical states $|\psi >$ are those satisfy the condition

\begin{equation}
L_{n}|\psi >=0\text{ }for\text{ }n>0,\text{ }L_{0}|\psi >=|\psi >.  \tag{2.9}
\label{Ln}
\end{equation}%
The massless tachyon%
\begin{equation}
Q_{j}=e^{i\beta _{j}X+\alpha _{j}\phi }  \tag{2.10}
\end{equation}%
are positive-norm physical states if either of the on-shell condition%
\begin{equation}
\pm (\beta _{j}-Q_{M})=(\alpha _{j}+Q_{L})  \tag{2.11}
\end{equation}%
is satisfied. If one defines%
\begin{equation}
\beta _{j}=jQ+(p-q)Q,  \tag{2.12}
\end{equation}%
then%
\begin{equation}
\alpha _{j}^{\pm }=\pm \left\vert jQ\right\vert -(p+q)Q.  \tag{2.13}
\end{equation}%
It's now easy to see that%
\begin{equation*}
Q_{2q}^{+}=e^{i(p+q)QX-(p-q)Q\phi },
\end{equation*}%
\begin{equation}
Q_{-2q}^{+}=e^{-i(p+q)QX+(p-q)Q\phi }  \tag{2.14}
\end{equation}%
together with%
\begin{equation}
\int \frac{dz}{2\pi i}O_{-2p}^{+}(z)O_{2q}^{+}(0)\sim \partial
_{z}[i(p+q)QX-(p-q)Q\phi ],  \tag{2.15}  \label{phy}
\end{equation}%
are the zero modes of generators of the level one $SU(2)_{k=1}$ Kac-Moody
algebra. Note that there is no concept of \textquotedblright material
gauge\textquotedblright\ as one has in the $c=1$ theory and the Liouville
field $\phi $ appears in (\ref{phy}). In general there exist discrete states
( $j=\{0,\frac{1}{2},1,...\}$ and $M=\{-J,-J+1,...,J\}$ )%
\begin{equation}
\Psi _{J,M}^{\pm }\sim \left( \int \frac{dz}{2\pi i}O_{-2p}^{+}(z)\right)
^{J-M}O_{2Jq}^{\pm }(0).  \tag{2.16}  \label{persy}
\end{equation}%
One can express the discrete states in (\ref{persy}) in terms of Schur
polynomials, which are defined as follows:%
\begin{equation}
\exp \left( \overset{\infty }{\underset{k=1}{\sum }}a_{k}x^{k}\right) =%
\overset{\infty }{\underset{k=1}{\sum }}S_{k}(a_{k})x^{k},  \tag{2.17}
\end{equation}%
where $S^{k}$ is the Schur polynomial, a function of $\{a_{k}\}=\{a_{i}:i\in
Z_{k}\}$. An explicit calculation of (\ref{persy}) gives 
\begin{eqnarray}
\Psi _{J,M}^{\pm } &\sim &\left\vert 
\begin{array}{cccc}
S_{2J-1} & S_{2J-2} & 
\begin{array}{ccc}
. & . & .%
\end{array}
& S_{J+M} \\ 
S_{2J-2} & S_{2J-3} & 
\begin{array}{ccc}
. & . & .%
\end{array}
& S_{J+M-1} \\ 
\begin{array}{c}
. \\ 
. \\ 
.%
\end{array}
& 
\begin{array}{c}
. \\ 
. \\ 
.%
\end{array}
& 
\begin{array}{ccc}
. &  &  \\ 
& . &  \\ 
&  & .%
\end{array}
& 
\begin{array}{c}
. \\ 
. \\ 
.%
\end{array}
\\ 
S_{J+M} & S_{J+M-1} & 
\begin{array}{ccc}
. & . & .%
\end{array}
& S_{2M+1}%
\end{array}%
\right\vert   \TCItag{2.18}  \label{Mpersy} \\
&&\times \exp \{i[(\pm J+M-1)q-(\pm J-M-1)q]QX  \notag \\
&&+[(\pm J+M-1)q+(\pm J-M-1)p]Q\phi \}  \notag
\end{eqnarray}%
with $S_{k}=S_{k}(\{\frac{1}{k!}\partial ^{k}[-iQ_{L}X+Q_{M}\phi ]\})$ and $%
S_{k}=0$ if \ $k<0$. We will denote the determinant in (2.18) $\Delta
(L,M,-iQ_{L}X+Q_{M}\phi )$.

In the OCQ of the theory, in addition to the positive-norm physical states
as discussed above, we still have an infinite number of $continuum$ $momentum
$ gauge states in the spectrum. They are solutions of either of the
following equations: 
\begin{equation}
|\psi >=L_{-1}|\chi >\text{ where }L_{m}|\chi >=0,\text{ }m\geq 0; 
\tag{2.19}  \label{persy1}
\end{equation}%
\begin{eqnarray}
|\psi  &>&=(L_{-2}+\frac{3}{2}L_{-1}^{2})|\xi >  \TCItag{2.20}
\label{persy2} \\
\text{ where }L_{m}|\xi  &>&=0\text{ , }m>0,\text{ (}L_{0}+1)|\xi >=0. 
\notag
\end{eqnarray}%
They satisfy the physical states conditions (\ref{Ln}), and have zero-norm.
Note that (2.20) is a gauge state only when the critical condition (\ref%
{CMCL}) is satisfied. It's difficult to give the general formula for all the
solutions of (\ref{persy1}) and (2.20). However, as was motivated from the $%
c=1$ theory \cite{T.D. Chung1}, we propose the following DGS for the $\Psi
^{-}$ sector 
\begin{eqnarray}
G_{J,M}^{-} &\sim &\left[ \int \frac{dz}{2\pi i}O_{0}(z)\right] \Psi
_{J-1,M}^{-}(0)  \notag \\
&\sim &\left[ \int \frac{dz}{2\pi i}e^{-i(p-q)QX-(p+q)Q\phi }(z)\right] \Psi
_{J-1,M}^{-}(0)  \notag \\
&\sim &S_{2J-1}(\{\frac{-1}{k!}\partial ^{k}[(p+q)Q\phi +i(p-q)QX]\})  \notag
\\
&&\Delta (J-1,M,-iQ_{L}X+Q_{M}\phi )  \notag \\
&&\exp \{i[(-J+M-1)q+(J+M+1)p]QX  \notag \\
&&+[(-J+M-1)q-(J+M+1)p]Q\phi \}.  \TCItag{2.21}  \label{GJM}
\end{eqnarray}%
It can be proved that these states are zero-norm states and satisfy the
physical states condition (\ref{Ln}). As an example, for $J=\frac{3}{2}%
,M=\pm \frac{1}{2},$%
\begin{eqnarray}
G_{\frac{3}{2},\pm \frac{1}{2}}^{-} &=&[Q_{L}^{2}(\partial \phi
)^{2}-2iQ_{L}Q_{M}\partial \phi \partial X-Q_{M}^{2}(\partial X)^{2}  \notag
\\
&&-Q_{L}\partial ^{2}\phi +iQ_{M}\partial ^{2}X]  \notag \\
&&\times e^{(\mp \frac{1}{2}Q_{M}-\frac{5}{2}Q_{L})\phi +(\pm \frac{i}{2}%
Q_{L}+i\frac{5}{2}Q_{M})X},  \TCItag{2.22}
\end{eqnarray}%
which can be shown to be mixture of solutions of (\ref{persy1}) and (2.20).
For the $\Psi ^{+}$ sector, we can subtract two positive-norm discrete
states as was done in the c = 1 theory \cite{T.D. Chung1} to obtain a gauge
state:%
\begin{equation}
G_{J,M}^{+}\sim \int \frac{dz}{2\pi i}[\Psi _{1,-1}^{+}(z)\Psi
_{J,M+1}^{+}(0)+\Psi _{J,M+1}^{+}(z)\Psi _{1,-1}^{+}(0)].  \tag{2.23}
\label{GJM2}
\end{equation}%
As an example, we have 
\begin{eqnarray}
G_{\frac{3}{2},\pm \frac{1}{2}}^{-} &=&%
\begin{array}{c}
\lbrack \frac{1}{4}\left( 
\begin{array}{cc}
-Q_{L}^{2}+5Q_{M}^{2}\pm 6iQ_{L}Q_{M} & \pm 3Q_{L}^{2}\pm
3Q_{M}^{2}-4iQ_{L}Q_{M} \\ 
\pm 3Q_{L}^{2}\pm 3Q_{M}^{2}-4iQ_{L}Q_{M} & -5Q_{L}^{2}+Q_{M}^{2}\mp
6iQ_{L}Q_{M}%
\end{array}%
\right)  \\ 
\times \partial X^{\mu }\partial X^{\nu }-i\frac{1}{2}\left( 
\begin{array}{c}
Q_{L}\mp 5iQ_{M} \\ 
-iQ_{M}\mp 5Q_{L}%
\end{array}%
\right) \partial X^{\mu }]%
\end{array}
\notag \\
&&\times e^{\frac{1}{2\sqrt{2}}(\mp Q_{M}+Q_{L})\phi +\frac{1}{2\sqrt{2}}%
(\pm iQ_{L}-iQ_{M})X},  \TCItag{2.24}
\end{eqnarray}%
which can be shown to be a solution of (2.20).

\section{SO(2,C) invariant and $w_{\infty }$ charges}

It can be easily seen that (2.18), (2.21) and (\ref{GJM2}) reduce to similar
equations in $c=1$ theory when we take the $p=q+1,q=$ $\infty $ limit. On
the other hand, the DGS in (2.21) and (\ref{GJM2}) can be obtained by doing
a $SO(2,C)$ rotation on that of the $c=1$ theory%
\begin{equation}
M=\frac{Q}{\sqrt{2}}\left( 
\begin{array}{cc}
p+q & i(q-p) \\ 
i(q-p) & q+p%
\end{array}%
\right) =\frac{1}{\sqrt{2}}\left( 
\begin{array}{cc}
Q_{L} & -iQ_{M} \\ 
iQ_{M} & Q_{L}%
\end{array}%
\right) ,  \tag{3.1}
\end{equation}%
where $M\in SO(2,C).$ A similar result was noticed in the ring structure of
BRST cohomology in \cite{B.Lian}. The $SO(2,C)$ invariant structure of the
DGS in the OCQ scheme suggests that the gauge state equations (\ref{persy1})
and (2.20) may possess a hidden symmetry property which has implication on
all the continuum momentum gauge states. Our results in this paper show once
again that the structure of DGS in the OCQ scheme is closely related to the
nontrivial ghost sector in the BRST quantization of the theory. Finally,
because of $SO(2,C)$ invariance, from (\ref{GJM2}) and a similar argument to
the $c=1$ theory, one can easily see that the $c<1$ DGS constructed here
also carrys the $w_{\infty }$ charges.

$Acknowledgements.$ This research is supported by National Science Council
of Taiwan, R.O.C., under grant number NSC852112-M-009-019

\end{document}